\documentclass[prd,twocolumn,longbibliography,nofootinbib,superscriptaddress,floatfix,preprintnumbers,hyperref,nofootinbib,nobibnotes,amsmath,amssymb,aps,preprintnumbers]{revtex4-1}

\usepackage{amsmath,amssymb,amsthm}
\usepackage{graphicx}
\usepackage[usenames, dvipsnames]{xcolor}
\usepackage{slashed}
\usepackage{subfig}
\usepackage{tabularx}
\usepackage{empheq}
\usepackage{mathrsfs}
\usepackage{comment}
\usepackage{bbold}
\usepackage[shortlabels]{enumitem}
\usepackage[normalem]{ulem}
\usepackage{url}
\usepackage{color}
\usepackage{soul}
\usepackage{appendix}

\definecolor{light-gray}{gray}{0.9}
\definecolor{medium-gray}{gray}{0.7}
\definecolor{darkblue}{rgb}{0.0,0.0,0.6}
\definecolor{red}{rgb}{0.9, 0,0}
\definecolor{navy}{rgb}{0.05, 0.05,0.8}
\definecolor{linkcolor}{rgb}{0.0,0.5,0.4}
\definecolor{paleblue}{rgb}{0.69, 0.93, 0.93}  
\definecolor{brown}{rgb}{.7,.35,.1}

\usepackage[colorlinks]{hyperref}
\hypersetup{
     colorlinks   = true,
     citecolor    = linkcolor,
     linkcolor    = linkcolor,
     urlcolor    = linkcolor
}

\usepackage{subcaption}
\usepackage{dcolumn}
\usepackage{bm}
\usepackage[mathlines]{lineno}
\usepackage{verbatim}
\usepackage{soul}
\usepackage[capitalise]{cleveref}

\DeclareUnicodeCharacter{2212}{\textendash}
\newcommand{\be}{\begin{equation}}
\newcommand{\ee}{\end{equation}}
\newcommand{\nl}{\nonumber \\}

\newcommand{\Ap}{A^\prime}
\newcommand{\Apm}{A^{\p \, \mu}}
\newcommand{\mAp}{m_{A^\prime}}
\newcommand{\Lag}{\mathscr{L}}

\newcommand{\mpl}{m_\text{pl}}
\newcommand{\p}{\prime}

\newcommand{\grad}{\nabla}

\newcommand{\order}[1]{\mathcal{O}{(#1)}}
\newcommand{\Eq}[1]{Eq.~\ref{eq:#1}} 

\newcommand{\Eqs}[2]{Eqs.~\ref{eq:#1} and \ref{eq:#2}}

\newcommand{\Fig}[1]{Fig.~\ref{fig:#1}}

\newcommand\mat[1]{\begin{pmatrix}#1\end{pmatrix}} 
\newcommand{\bebox}{\begin{empheq}[box=\fcolorbox{light-gray}{light-gray}]{align}}
\newcommand{\eebox}{\end{empheq}}

\newcommand{\rhodm}{\rho_{_\text{DM}}}
\newcommand{\DM}{{_\text{DM}}}

\newcommand{\dt}{\partial_t}

\newcommand{\ket}[1]{|#1 \rangle}
\newcommand{\bra}[1]{\langle #1 |}

\newcommand{\eqbox}[1]{\text{\fcolorbox{light-gray}{light-gray}{$#1$}}}

\newcommand{\w}{\omega}

\newcommand{\n}{\nu}
\newcommand{\nbar}{\bar{\nu}}
\newcommand{\g}{\gamma}

\newcommand{\vv}{{\bf v}}
\newcommand{\Vv}{{\bf V}}
\newcommand{\Av}{{\bf A}}
\newcommand{\Bv}{{\bf B}}

\newcommand{\Fv}{{\bf F}}

\newcommand{\av}{\boldsymbol{a}}

\newcommand{\sv}{{\bf s}}
\newcommand{\Sv}{{\bf S}}

\newcommand{\Omegav}{{\bf \Omega}}

\newcommand{\nv}{\boldsymbol{n}}

\newcommand{\sigmav}{\boldsymbol{\sigma}}

\newcommand{\eV}{\text{eV}}
\newcommand{\meV}{\text{meV}}

\newcommand{\MeV}{\text{MeV}}
\newcommand{\GeV}{\text{GeV}}
\newcommand{\TeV}{\text{TeV}}

\newcommand{\cm}{\text{cm}}

\newcommand{\s}{\text{s}}

\newcommand{\AU}{\text{AU}}

\newcommand{\kpc}{\text{kpc}}

\newcommand{\Uone}{\text{U}(1)}

\newcommand{\Boron}{{}^8\text{B}}
\newcommand{\nnbar}{\n \leftrightarrow \nbar}

\newcommand{\vp}{\mathbf{p}}

\newcommand{\Prn}{P_{\n \nbar}}
\newcommand{\shat}{\hat{\sv}}
\newcommand{\nrf}{\n \text{RF}}
\newcommand{\wv}{\boldsymbol{\omega}}
\newcommand{\fv}{\boldsymbol{f}}

\definecolor{brightpink}{rgb}{1.0, 0.0, 0.5}
\definecolor{emerald}{rgb}{0.31, 0.78, 0.47}
\definecolor{limegreen}{rgb}{0.2, 0.8, 0.2}
\definecolor{blue-violet}{rgb}{0.33, 0.17, 0.89}

\newcommand{\downket}{| \hspace{-0.1cm} \downarrow \rangle}
\newcommand{\upket}{| \hspace{-0.1cm} \uparrow \rangle}

\newcommand{\upbra}{\langle \uparrow \hspace{-0.1cm} |}

\begin{document}

\preprint{FERMILAB-PUB-26-0081-SQMS-T}


\title{Neutrino-Antineutrino Conversion from Ultralight Vector Dark Matter}

\author{Asher Berlin}
\affiliation{Theoretical Physics Division, Fermi National Accelerator Laboratory, Batavia, IL 60510, USA}
\affiliation{Superconducting Quantum Materials and Systems Center (SQMS), Fermi National Accelerator Laboratory}

\author{Rodolfo Capdevilla}
\affiliation{Theoretical Physics Division, Fermi National Accelerator Laboratory, Batavia, IL 60510, USA}
\affiliation{Atlantis University, Miami, FL 33136, USA}

\author{Ting Cheng}
\affiliation{Theoretical Physics Division, Fermi National Accelerator Laboratory, Batavia, IL 60510, USA}

\author{Matheus Hostert}
\affiliation{Department of Physics and Astronomy, University of Iowa, Iowa City, IA 52242, USA}

\author{Pedro A. N. Machado}
\affiliation{Theoretical Physics Division, Fermi National Accelerator Laboratory, Batavia, IL 60510, USA}


\begin{abstract}
We show that Majorana neutrinos convert into antineutrinos in a background of ultralight vector dark matter coupled to lepton number, such as the gauge boson of $\text{U}(1)_{B-L}$ or $\text{U}(1)_{L_i - L_j}$ with $i, j = e , \mu, \tau$. This effect is suppressed by the small neutrino mass, but the enhancement by long astrophysical baselines can enable future searches for solar and supernova neutrinos to explore uncharted parameter space. For instance, for $\text{U}(1)_{B-L}$ dark matter, the observation of a supernova neutrino burst at DUNE, Hyper-Kamiokande, and JUNO could probe gauge couplings as small as $e^\prime \sim 10^{-32} - 10^{-25}$ for dark matter masses of $m_{A^\prime} \sim 10^{-22} \ \text{eV} - 10^{-14} \ \text{eV}$, beyond the capability of other future probes.
\end{abstract}


\maketitle


The smallness of the observed neutrino masses is naturally explained by the inclusion of Majorana mass terms that break lepton number~\cite{Minkowski:1977sc,Weinberg:1979sa,Gell-Mann:1979vob,Mohapatra:1979ia,Yanagida:1980xy,Schechter:1980gr,Schechter:1981cv,Lazarides:1980nt,Mohapatra:1980yp,Foot:1988aq}. If lepton number is part of a new local symmetry that is preserved in the ultraviolet, but broken spontaneously at low energies, then there exists a new massive gauge boson $\Apm$ coupled to neutrinos. Notable anomaly-free examples include the gauge bosons of  $\Uone_{B-L}$ and  $\Uone_{L_i - L_j}$, where $i, j = e , \mu , \tau$. Such interactions can also arise in theories where the spontaneous breaking of a secluded $\Uone^\p$ results in mass-mixing between active neutrinos and new dark-charged fermions~\cite{Fox:2011qd,Schmaltz:2017oov}. 

Vectors are naturally light, and thus constitute compelling dark matter (DM) candidates across a wide range of masses. However, most studies of neutrino-coupled vector DM have only investigated a  subset of possible signals. For instance, past work has dominantly focused on modifications to flavor oscillations~\cite{Brzeminski:2022rkf}, with only a few studies investigating spin-related signals from backgrounds of ultralight vectors (see, e.g., Refs.~\cite{Janish:2020knz,Fang:2025fvo}). More generally, signatures of neutrino-coupled sub-eV DM have been investigated before, but past studies have mainly explored effects that can only arise from fine-tuned scalar DM, such as corrections to neutrino masses~\cite{Berlin:2016woy, Krnjaic:2017zlz, Brdar:2017kbt, Capozzi:2018bps, Kelly:2019wow, Dev:2020kgz, Losada:2021bxx, Huang:2021kam, Dev:2022bae, Losada:2022uvr, Losada:2023zap, Gherghetta:2023myo}. 

In this work, we examine experimental signatures arising from the coupling of ultralight vector DM to the spin of a Majorana neutrino, which can cause the helicity of astrophysical neutrinos to flip at a detectable level. In the relativistic limit, this is observationally equivalent to neutrinos $\n$ oscillating with antineutrinos $\nbar$, with the necessary lepton-number violation provided by the Majorana mass. Since any such $\Ap$ directly or radiatively couples to electrons and nucleons, strong bounds on these models are typically derived from fifth-force searches. Regardless, as we show below, astrophysical signals of $\nnbar$ conversion are enhanced by long  baselines, enabling sensitivity to parameter space orders of magnitude beyond existing constraints. We use this to show how future measurements of solar and Galactic supernova (SN) neutrinos by upcoming experiments, such as DUNE, Hyper-Kamiokande, and JUNO, can enable vast new regions of parameter space to be explored for vector DM, including $\Uone_{B-L}$.

\vspace{0.25cm}
\noindent \textbf{Overview.}---
If $\Ap$ is an ultralight coherent field and accounts for the local DM density $\rhodm \simeq 0.4 \ \GeV / \cm^3$, then its large phase-space occupancy implies that it behaves as a classical non-relativistic field. In the laboratory frame, such a field oscillates at a frequency given by its  mass $\mAp$ and with a spatial gradient set by the speed of the DM wind $v_\DM \sim 10^{-3}$. Within a DM coherence length, \mbox{$\lambda_\DM \simeq 2 \pi / (\mAp \, v_\DM)$}, the spatial components of $\Apm$ can be written in terms of the  orthonormal Cartesian basis vectors $\hat{\nv}_i$ as~\cite{Fedderke:2021aqo,Amaral:2024tjg}
\be
\label{eq:DMfield}
\Av^\p (t) \simeq \frac{\sqrt{2 \rhodm/3}}{\mAp} \, \text{Re} \Big[ e^{i \mAp t} \, \sum_{i = 1}^3 \alpha_i \, e^{i \phi_i} \, \hat{\nv}_i \Big]
~.
\ee
Above, $\alpha_i$ are $\order{1}$ Rayleigh-distributed variables, and $\phi_i $ are uniformly-distributed phases, both of which vary randomly over distance scales comparable to $\lambda_\DM$. Note that \Eq{DMfield} corresponds to an elliptically polarized $\Av^\p$ field, since it can generically be decomposed into two components with comparable magnitude but opposite phase. Thus, $\Av^\p$ traces out a randomly-oriented ellipse at frequency $\mAp$.  We will largely ignore the time-component of $\Apm$, since it is suppressed by the DM velocity, \mbox{$A^{\p \, 0} \simeq \vv_\DM \hspace{-0.05 cm} \cdot \Av^\p$}, as required by conservation of the dark current density $j^{\p \, \mu}$ for a massive Proca field, $\partial_\mu \Apm \propto \partial_\mu j^{\p \, \mu} = 0$~\cite{Berlin:2024pzi} (up to negligible corrections from Majorana neutrino masses).

We focus specifically on DM interactions with the spin of a neutrino. For a single Majorana neutrino generation of mass $m_\n$, the relevant low-energy Lagrangian density, written in two-component fermion notation, is
\be
\label{eq:Lag1}
\Lag \supset e^\p \, \Ap_\mu \, \n^\dagger \bar{\sigma}^\mu \n - \bigg( \frac{1}{2} \, m_\n \, \n \n + \text{h.c.} \bigg)
~,
\ee
where $e^\p \ll 1$ parametrizes the strength of the $\Ap$-$\n$ interaction.

It is simple to discern the resulting physical effects of \Eq{Lag1} at a heuristic level. In particular, we see that $\Ap$ couples to the axial neutrino current. For a definite-helicity neutrino population of number density $n_\n$, velocity $\vv$, and spin-direction $\shat$, the classical analogue of the axial-current is the covariant spin-density four-vector \mbox{$\langle \n^\dagger \bar{\sigma}^\mu \n \rangle \to n_\n \, (\vv \cdot \shat \, , \, \shat)^\mu$}~\cite{Jackson:1998nia}. Thus, the interaction term in \Eq{Lag1} yields the single-particle classical Lagrangian
\be
\label{eq:ClassicalLag1}
L \supset (q \, \Av)_\text{eff} \cdot \vv + (\mu \, \Bv)_\text{eff} \cdot \shat
~,
\ee
where we defined the effective vector potential \mbox{$(q \, \Av)_\text{eff}  = - e^\p \, A^{\p \, 0} \, \shat$} and effective magnetic dipole interaction \mbox{$(\mu \, \Bv)_\text{eff} = e^\p \, \Av^\p$}. 

Since $(q \, \Av)_\text{eff}$ couples to the neutrino velocity, it gives rise to a force analogous to how an electromagnetic vector potential enters the Lorentz force law,
\be
\label{eq:force1}
\Fv = e^\p \Big( \dt A^{\p \, 0} \, \shat + \vv \times \big( \shat \times \grad A^{\p \, 0} \big) \Big)
~.
\ee
Note that this can also be derived directly from the Heisenberg equations of motion~\cite{Berlin:2023ubt}.

The second term of \Eq{ClassicalLag1} couples $(\mu \, \Bv)_\text{eff}$ to the spin. As a result, the neutrino spin $\shat$ precesses around $\Av^\p$ in the neutrino rest frame ($\nrf$). In the lab-frame, the neutrino spin therefore precesses at a rate of
\be
\label{eq:precess1}
\dt \shat = - \frac{e^\p}{\g} \, \Av_{\nrf}^\p \times \shat + \wv_T \times \shat
~,
\ee
where $\Av^\p_{\nrf}$ is the $\Av^\p$ field evaluated in the $\nrf$, and $\wv_T =  \frac{\g^2}{2 (\g+1)} \, (\vv \times \av)$ incorporates the effect of Thomas precession for a neutrino with acceleration $\av$ and boost $\g$. In our case, the only force that can contribute to $\wv_T$ is the acceleration generated by the second term of \Eq{force1}. However, the resulting contribution to Thomas precession is negligible compared to the first term of \Eq{precess1} provided that $\mAp \ll m_\n / (\g \, v_\DM^2)$, which is satisfied throughout the parameter space of interest. We thus focus on the first term for the remainder of this work.

A more detailed derivation of these effects is given in the Supplemental Material. There, it is shown that the interaction of \Eq{Lag1} modifies the quantum mechanical Hamiltonian of a neutrino, which in its rest frame is given by
\be
H \supset - e^\p \, \Av^\p_{\nrf} \cdot \sigmav
~,
\ee
where $\sigmav$ is the vector of Pauli matrices acting on the neutrino spin wavefunction. This again shows that the interaction is analogous to that of a magnetic dipole, in the sense that it splits the neutrino spin's energy levels and thus induces spin-precession  around $\Av^\p$. Such effects are well-studied within the context of electrons or nucleons coupled to axion DM~\cite{Krauss:1985ub,Raffelt:1985pvi,Slonczewski:1985oco,Graham:2017ivz,Berlin:2023ubt}. In this case, the precession of a polarized macroscopic sample of electron/nucleon spins can give rise to bulk mechanical or electromagnetic signals. 

What is the analogous effect for a neutrino? The precession of a relativistic Majorana neutrino corresponds to a flip in helicity, while leaving chirality unchanged. At the level of experimental observables, this is therefore equivalent to $\nnbar$ conversion (throughout the remainder of this work, we will  use ``negative-helicity neutrinos" and ``neutrinos" interchangeably, as well as ``positive-helicity neutrinos" and ``antineutrinos"). This is shown explicitly at the level of quantum field theory in the Supplemental Material.

Hence, in a DM background the interactions of \Eq{Lag1} can give rise to anomalous antineutrino-like events from astrophysical sources that are otherwise expected to produce mainly neutrinos. As shown below, the rate for $\nnbar$ conversion is suppressed by the large neutrino boost but enhanced by the long baseline between the source and the Earth. DM-induced neutrino precession is similar to the effect of a neutrino's magnetic moment in a background magnetic field~\cite{Okun:1986na,Raffelt:1996wa}. However, unlike a magnetic moment, the different Lorentz structure in \Eq{Lag1} means that our signal applies even to a single generation of Majorana neutrinos in the presence of a DM background. 

\vspace{0.25cm}
\noindent \textbf{Neutrino-Antineutrino Conversion.}---
The $\nnbar$ conversion probability can be derived most straightforwardly from unitary time-evolution of an initially relativistic neutrino with definite helicity traveling for a distance $L$.  In the Supplemental Material, we write down a representative Hamiltonian corresponding to the neutrino spin's coupling to an elliptically-polarized $\Av^\p$ field. As shown there, it is possible to find an approximate analytic form for the $\nnbar$ transition probability $\Prn$ in certain limits. In particular, for $\mAp^2 / (e^\p \sqrt{\rhodm}) \gg 1$ and $L \geq \lambda_\DM$, we find, up to $\order{1}$ factors,
\be
\label{eq:PrEllip1}
\Prn \sim \frac{L}{\lambda_\DM} \, \bigg( \frac{e^{\p \, 2} \rhodm}{\mAp^3} \,  \, \frac{\lambda_\DM}{\g} \bigg)^2
~.
\ee
On the other hand, for $1 / \g^2 \ll \mAp^2 / (e^\p \sqrt{\rhodm}) \ll 1 $,
\be
\label{eq:PrEllip2}
\Prn \sim \frac{e^\p \sqrt{\rhodm} \, L}{\g^2} \, \max\big( \mAp^{-1} ~,~ T \big)
~,
\ee
where we defined the timescale 
\be
T \equiv \min\bigg( L ~,~ \frac{\mAp^2 \, \lambda_\DM}{e^\p \sqrt{\rhodm}}  \bigg)
~.
\ee
We have also confirmed numerically that the above expressions are accurate for $\Prn \lesssim 1$ and saturate for $\Prn \sim \order{1}$. This shows that the conversion probability is enhanced for larger $L$ and smaller $\g$. We are therefore motivated to consider distant sources that emit low-energy neutrinos with an asymmetric helicity. 

Before proceeding, let us comment on some effects that can, in principle, modify \Eqs{PrEllip1}{PrEllip2}. For instance,  the density of normal matter $n_\text{SM}$ contributes to the $\nrf$ Hamiltonian as \mbox{$H_\text{SM} \sim \g \, (G_F \, n_\text{SM} + e^\p \, A^{\p \, 0}_\text{SM} ) \,  \vv \cdot \sigmav$}, where $G_F$ is Fermi's constant and $\Ap_\text{SM}$ is the lab-frame $\Ap$ field sourced by the density of SM particles. These contributions do not efficiently induce neutrino spin-precession, since $\vv$  is dominantly aligned with the neutrino spin. However, as discussed in the Supplemental Material, if \mbox{$H_\text{SM} \gg \max{(e^\p \, \Ap_\text{DM} \, , \, e^\p \, \Ap_\text{DM} \, \g \, v_\DM \, , \, \g \, \mAp \, v_\DM)}$} with \mbox{$\Ap_\text{DM} \sim \sqrt{\rhodm} / \mAp$}, then $\Prn$ is substantially reduced. This is not a concern for the parameter space of interest in this work.

We also note that in deriving \Eq{PrEllip1}, we considered interactions with only a single neutrino mass-eigenstate. This therefore applies straightforwardly to models where $\Ap$ couples diagonally to the mass-eigenstates $\n_{1,2,3}$, which is the case for $\Uone_{B-L}$. However, in flavor-dependent theories, such as $\Uone_{L_i - L_j}$, off-diagonal interactions are also generated. For instance, due to the form of the PMNS matrix, the gauge boson of $\Uone_{L_\mu - L_\tau}$ has unsuppressed off-diagonal interactions with $\nu_{1,2}$--$\nu_3$ and mass-diagonal interactions that are relatively suppressed by $\order{0.1}$. Such off-diagonal interactions can lead to processes of the form $\n_1 \leftrightarrow \bar{\n}_3$, etc., but the corresponding rate is suppressed by the neutrino mass-splitting \mbox{$(e^\p |\Av^\p|)^2 / (m_3 - m_1)^2 \ll 1$}. In the remainder of this study, we therefore focus strictly on mass-diagonal interactions, as in \Eq{Lag1}.

\vspace{0.25cm}
\noindent \textbf{Solar Signals.}---
Solar neutrinos are sensitive to DM-induced $\nnbar$ conversion. In particular, solar neutrinos from the decay $\Boron \to {}^8\text{Be}^* \, e^+ \, \n_e$ are produced in large numbers with a total flux at the Earth of $\Phi_\n \sim 6 \times 10^6 \ \cm^{-2} \ \s^{-1}$ and average energy of $ E_\n \simeq 7 \ \MeV$~\cite{Bahcall:1996qv}. If a fraction of these neutrinos convert to antineutrinos, they can be detected through inverse beta decay (IBD), $\nbar_e  \, p \to e^+ n$, in terrestrial detectors. This is a clean experimental signature exclusive to antineutrinos, based on the double-coincidence of a prompt positron signal and delayed neutron capture. Since the Standard Solar Model predicts a negligible flux of antineutrinos with energy above $\sim 3 \ \MeV$~\cite{Malaney:1989hs,Vitagliano:2017odj}, the absence of IBD signals at this energy successfully constrains a variety of new physics scenarios~\cite{Lim:1987tk,Akhmedov:1988uk,Berezhiani:1987gf,Berezhiani:1991vk,Hostert:2020oui,Picoreti:2021yct,deGouvea:2023jxn,Wu:2023twu,Ansarifard:2024zxm}. 

Experimental limits on the solar antineutrino flux have been set by the liquid scintillator detectors Borexino~\cite{Borexino:2019wln} and KamLAND~\cite{KamLAND:2011bnd,KamLAND:2021gvi}, as well as by Super-Kamiokande, a Gd-doped water Cherenkov detector~\cite{Super-Kamiokande:2020frs,Super-Kamiokande:2023xup}. KamLAND provides some of the strongest limits on $\nnbar$ conversion of $\Boron$ neutrinos, bounding $\Prn \lesssim 3.5 \times 10^{-5}$ under the assumption that the conversion probability is energy-independent~\cite{KamLAND:2021gvi}. This is derived from observations spanning multiple energy bins, but is mostly driven by the upper limit of $\sim 10 \ \cm^{-2} \ \s^{-1} \ \MeV^{-1}$ on the differential $\nbar$ flux at an energy of $\sim 10 \ \MeV$. 

Due to the Mikheyev-Smirnov-Wolfenstein (MSW) effect, over $90\%$ of $\Boron$ neutrinos exit the Sun's interior as the mass-eigenstate $\n_2$~\cite{Nunokawa:2006ms}. 
Therefore, solar antineutrino searches are primarily sensitive to the conversion $\n_2 \to \nbar_2$, which occurs with a probability given by the expressions provided in the previous section, with $m_\n = m_2$, $L = 1 \ \AU$, and an additional factor of $\cos^2\theta_{13} \, \sin^2{\theta_{12}} \simeq 0.3$ to account for the likelihood of detecting $\nbar_2$ through IBD.

Unlike an energy-independent conversion probability, DM-induced $\nnbar$ transitions are suppressed by the neutrino's boost, and are thus peaked at small energy. However, sensitivity to lower energy conversions is parametrically weakened both by the falling $\Boron$ flux and increased backgrounds. We have calculated the differential flux of $\nbar_e$ using the $\Boron$ spectrum of Ref.~\cite{Bahcall:1996qv} and compared this to bounds from Refs.~\cite{Borexino:2019wln,KamLAND:2011bnd,KamLAND:2021gvi,Super-Kamiokande:2020frs,Super-Kamiokande:2023xup}, which shows that DM-induced conversion is still mainly constrained by energies near $\sim 10 \ \MeV$. The latest KamLAND search~\cite{KamLAND:2021gvi} places the strongest limits, followed by limits from Super-Kamiokande and Borexino.

Future upgrades of existing detectors and new detectors coming online in the coming years are expected to increase this sensitivity substantially, due to larger exposures, lower thresholds, and reduced backgrounds~\cite{Hostert:2020oui}. For instance, detectors such as JUNO~\cite{JUNO:2015zny} and Jinping~\cite{Jinping:2016iiq} are projected to obtain sensitivity to conversion probabilities of $\sim 10^{-6} - 10^{-5}$~\cite{Li:2019snw, Franklin:2023diy}. 
As a rough estimate, we rescale the previous KamLAND~\cite{KamLAND:2021gvi} limit from 6.7~kt-yr to 100~kt-yr assuming negligible backgrounds, corresponding to a sensitivity of $\Prn \sim 2 \times 10^{-6}$.

\vspace{0.25cm}
\noindent \textbf{Supernova Signals.}---
Observations of neutrinos emitted from the core-collapse of a Galactic SN can also be used to constrain $\nnbar$ conversion. To understand this, let us first briefly review the distinct phases of SN neutrino emission~\cite{Mirizzi:2015eza,Scholberg:2017czd}. 

The neutronization burst occurs during the first $\sim 10 \ \text{ms}$ shock breakout phase, in which the core-collapse shockwave liberates protons from nuclei, which subsequently capture electrons, $p \, e^- \to n \, \n_e$, producing a sharp initial burst of $\n_e$ that dominates over the flux of other neutrino species. This is followed by the accretion phase, which lasts $\sim 100 \ \text{ms}$ and emits all neutrino flavors with distinct temperature hierarchies, $T_{\n_e} < T_{\nbar_e} < T_{\n_x}$ (where $\n_x$ denotes all other flavors), reflecting the flavor-dependent interaction rates in the proto-neutron star. Normal neutrino oscillations may non-trivially impact the energy spectra of different flavors~\cite{Fischer_2010, Serpico:2011ir, Mirizzi:2015eza, Scholberg:2017czd, Capozzi:2018rzl}. 
Nevertheless, within the standard three-neutrino framework there is no mechanism that leads to equilibration between $\n$ and $\nbar$~\cite{Volpe:2023met}. 
The distinct energy spectrum between $\n$ and $\nbar$  will therefore enable a future Galactic SN observation to detect efficient DM-induced conversions, since this  would yield unexpectedly similar energy distributions at terrestrial detectors. 

In the near future, the DUNE experiment will be capable of measuring the SN $\n_e$ spectrum via $\n_e  \, {}^{40}\text{Ar}\to e^- \, {}^{40}\text{K}^*$, while Hyper-Kamiokande and JUNO will be sensitive to SN $\nbar_e$ via  IBD~\cite{DUNE:2020zfm, DUNE:2023rtr, Hyper-Kamiokande:2018ofw, Hyper-Kamiokande:2021frf, JUNO:2023dnp}. Sensitivity to $\n\leftrightarrow \bar\n$ transitions may be achieved by then comparing the spectra measured in these experiments as a function of arrival time. We conservatively estimate that future SN observations will be sensitive to $\n\leftrightarrow \bar\n$ transitions if the probability for this to occur is $\Prn \sim \order{1}$. In this case, the spectra of $\n_e$ and $\nbar_e$ are identical, and are thus distinguishable from standard SN neutrino physics~\cite{Akhmedov:2003fu,Ando:2003is,deGouvea:2019goq,Kopp:2022cug,Jana:2022tsa,Huang:2023aob}.

Aside from equilibration of the neutrino and antineutrino spectra from the accretion phase, the neutronization burst offers another potential signal, since efficient conversions would also give rise to an initial $\nbar_e$ burst. While this would be a striking signature of $\n\to\nbar$ transitions~\cite{deGouvea:2019goq}, the magnitude of this neutronization burst signal depends on the neutrino mass ordering, which is not yet determined. MSW transitions within the SN envelope substantially modify the emergent neutrino flavor composition.\footnote{The $\Ap$ field directly sourced by SM matter can also contribute to the neutrino's matter potential, but  this is typically subdominant compared to the contribution from electroweak interactions.} For inverted ordering (IO), $m_3 < m_1 < m_2$, $\n_e$ predominantly exits as $\n_2$, such that the neutronization phase yields a substantial $\n_e$ flux at Earth, due to the large mixing angle $\sin^2{\theta_{12}} \simeq 0.3$. Instead, for normal ordering (NO),  $m_1 < m_2 < m_3$, $\nu_e$ exits as $\n_3$, and the $\n_e$ burst flux is strongly suppressed at Earth by $\sin^2\theta_{13} \simeq 0.02$. Note that self-induced neutrino flavor conversion can lead to flavor equilibration among $\n_e,\,\n_\mu,\, \n_\tau$ and among $\bar\n_e,\,\bar\n_\mu,\, \bar\n_\tau$, but the spectra of $\n_e$ and $\bar\n_e$ would still be different~\cite{Fischer_2010, Serpico:2011ir, Capozzi:2018rzl}. While the accretion phase signal would be sufficient to claim new physics, the possibility of a neutronization burst in antineutrinos would provide an excellent experimental cross-check of $\n \to \nbar$ transitions.
In addition to Hyper-Kamiokande and JUNO, an early $\nbar_e$ burst could also be observed by IceCube as an excess of IBD events above constant backgrounds~\cite{IceCube:2023ogt}.

\begin{figure}[t]
\centering
\includegraphics[width=0.5 \textwidth]{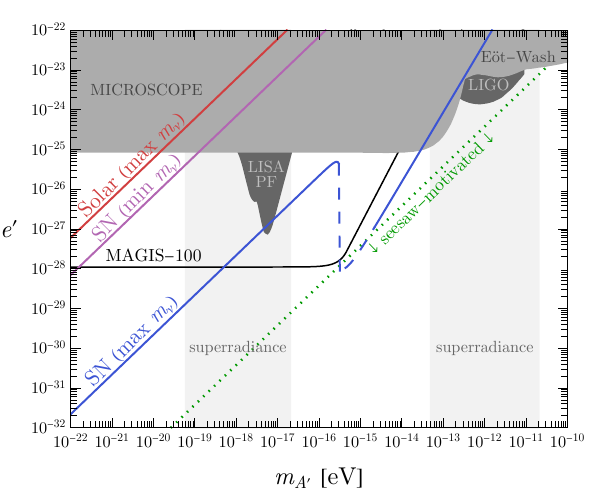}
\caption{The projected sensitivity of future solar (red) or supernova (purple and blue) neutrino experiments to $\n \to \nbar$ conversion induced by vector dark matter $\Ap$, as a function of its mass $\mAp$ and neutrino coupling $e^\p$. The solar line adopts the largest laboratory allowed neutrino mass, whereas the two SN lines correspond to a value for the relevant neutrino mass that is at the minimum or maximum of the experimentally allowed range. The solid parts of these lines rely on analytic approximations, whereas the dashed parts are an extrapolation of our analytic results. Also shown in gray are existing bounds~\cite{Wagner:2012ui,MICROSCOPE:2022doy,Frerick:2023xnf,LIGOScientific:2021ffg,Baryakhtar:2017ngi}. For lab-based bounds, we have taken the nucleon and electron coupling to be of the same magnitude as the neutrino coupling, which is the case for a model of $\Uone_{B-L}$. Aside from the superradiance bounds, these limits are suppressed when the couplings to matter particles are radiatively generated, as in models of $\Uone_{L_\mu - L_\tau}$ or where the neutrino interaction arises from mixing with a new sterile neutrino. The region of parameter space below the dotted green line is motivated by the seesaw generation of neutrino masses.}
\label{fig:reach}
\end{figure}

\vspace{0.25cm}
\noindent \textbf{Results.}---
The sensitivity of solar and SN neutrino experiments to vector DM is shown by the colored lines in \Fig{reach}. Also shown in gray are existing bounds from fifth-force searches by the E\"{o}t-Wash~\cite{Wagner:2012ui} and MICROSCOPE~\cite{MICROSCOPE:2022doy} collaborations, DM searches using the LISA Pathfinder~\cite{Frerick:2023xnf} and LIGO/Virgo interferometers~\cite{LIGOScientific:2021ffg}, as well as black hole superradiance~\cite{Arvanitaki:2009fg,Baryakhtar:2017ngi}. 
For the lab-based limits, we have taken the corrected ones as updated in Refs.~\cite{Amaral:2024tjg,Hamaide:2025buy}. 

For all but the superradiance limit, these existing bounds depend on the coupling of $\Ap$ to matter particles, such as nucleons and electrons. Here, we have conservatively assumed that the $\Ap$ couples to matter particles with the same magnitude as neutrinos, which is the case for $\Uone_{B-L}$. However, it is important to note that these bounds are significantly relaxed if the matter couplings are generated at the one-loop level, such that they are relatively suppressed by $\sim g^2 / 16 \pi^2$. This corresponds to theories where the first lepton generation is not directly charged under the $\Ap$, such as $\Uone_{L_\mu - L_\tau}$, or where the neutrino interaction is generated through the mixing with a new sterile neutrino. 

We find that past measurements of solar neutrinos do not place new bounds that are competitive with existing constraints. However, new parameter space can be explored by future solar neutrino measurements, provided that the absolute mass scale of the neutrino is near the largest laboratory-allowed value $m_\n \simeq 0.45 \ \eV$~\cite{KATRIN:2024cdt}, as illustrated by the red curve in \Fig{reach} which assumes an exposure of 100~kt-yr. 

Also shown in \Fig{reach} is the projected sensitivity of a SN observation by upcoming experiments such as DUNE, Hyper-Kamiokande, and JUNO, assuming either the minimum ($m_\n \simeq 8 \ \meV$) or maximum ($m_\n \simeq 0.45 \ \eV$)  allowed neutrino mass-scale for $\n_2$. Here, we have taken a neutrino energy of $E_\n = 10 \ \MeV$ and a SN distance of $L = 10 \ \kpc$. This shows that an observation of SN neutrinos could extend the sensitivity of, e.g., $\Uone_{B-L}$ vector DM by many orders of magnitude. This sensitivity to $\Uone_{B-L}$ DM is competitive with, and in certain regions of parameter space, outperforms that of other proposed experiments, such as the MAGIS-100 atom interferometer~\cite{MAGIS-100:2021etm}, which is shown as the black curve in \Fig{reach}.

There have also been bounds derived for new vectors that couple to  neutrinos with flavor-dependent interactions, such as $\Uone_{L_\mu - L_\tau}$. In this case, the spin-coupling of vector DM shifts the relative energies of different neutrino generations, modifying neutrino flavor oscillations~\cite{Wise:2018rnb,Coloma:2020gfv,Brzeminski:2022rkf,Arguelles:2024cjj}. In contrast, the signal discussed in our work does not rely on multiple generations and is present in models for which only a single flavor of neutrino couples to the $\Ap$ (or all neutrinos couple with equal strength). As a result, we do not display limits on such flavor-dependent interactions in \Fig{reach}. 

The region of parameter space below the dotted green line in \Fig{reach} is motivated by the seesaw mechanism for neutrino mass. In particular, we take the gauge boson mass to be $\mAp \sim e^\p v^\p$, where $v^\p$ is the VEV of a dark Higgs, and assume that there is a right-handed sterile neutrino $N$ with mass  $m_N \sim y_N \, v^\p$. After mixing with the active neutrino, we have $m_\n \sim (y_\n v)^2 / m_N$, where $v$ is the Standard Model Higgs VEV, which then implies \mbox{$e^\p \sim y_N \, m_\n \, \mAp / (y_\n v)^2$}. In \Fig{reach}, we show this parameter space after restricting the neutrino Yukawa $y_\n$ to be no smaller than the electron Yukawa in the Standard Model, $y_\n \gtrsim m_e / v$, as well as $y_N \lesssim 1$, which gives $e^\p \lesssim m_\n \, \mAp / m_e^2$.

Note that the introduction of $N$ renders $\Uone_{B-L}$ anomaly-free. However, we have implicitly assumed that its mass is sufficiently large ($m_N \gg 10 \ \MeV$) to be irrelevant to the phenomenology discussed above. In the absence of $N$, $\Uone_{B-L}$ has a cubic self-anomaly as well as a mixed gravitational one. Thus, decoupling $m_N$ to very high energies implies that the $\Ap$ has an anomalous radiative correction to its mass that enters at three loops (via the joining of two triangle-like diagrams), i.e., $\Delta \mAp \sim e^\p \, g_X^2 \, m_N / (4 \pi)^3$, where $g_X \sim \max{(e^\p , m_N / \mpl)}$ and $\mpl$ is the Planck mass~\cite{Gross:1972pv,Bouchiat:1972iq,Preskill:1990fr}. In the parameter space of interest, this is a negligible correction to $\mAp$ provided that $m_N \lesssim 10^{15} \ \GeV$. We also note that for $m_N \gg \TeV$, the explicit lepton-number breaking from the Majorana neutrino masses can induce processes (such as exotic $W$-boson decays) with rapidly energy-growing amplitudes; these are constrained by collider searches~\cite{Dror:2020fbh,Ekhterachian:2021rkx}, but the precise bounds depend on the UV-completion at scales higher than those we consider here.

In \Fig{reach}, we have restricted to DM masses greater than $10^{-22} \ \eV$, corresponding to de Broglie wavelengths smaller than the typical size of dwarf galaxies. However, detailed studies of stellar kinematics in ultra-faint dwarf galaxies have been used to place a stronger lower limit on the DM mass, corresponding to $\gtrsim 3 \times 10^{-19} \ \eV$~\cite{Dalal:2022rmp,May:2025ppj}. Regardless, such bounds typically do not apply to particles that make up a small fraction $f_\DM \ll 1$ of the DM density. In this case, our projections in \Fig{reach} can be extrapolated via $e^\p \propto \mAp / \sqrt{f_\DM}$ to  masses as small as $\mAp \sim (\g \, L)^{-1}$.

\vspace{0.25cm}
\noindent \textbf{Discussion.}---
In this work, we have shown that a background of ultralight vector DM can induce the precession of a neutrino's spin, which appears as $\nnbar$ conversion in terrestrial experiments. We have used this to show that upcoming solar and supernova neutrino experiments, such as DUNE, Hyper-Kamiokande, and JUNO, will be sensitive to orders of magnitude of previously unexplored parameter space for new ultralight vectors.

While many searches for vector DM directly rely on the local value of $\rhodm$, our effect depends on the integrated value between the neutrino source and the Earth. This is especially interesting in light of the fact that early universe production mechanisms for vector DM are often characterized by enhanced power on small scales~\cite{Graham:2015rva}. As a result, it is possible that  a significant fraction of such DM collapses into gravitationally bound solitons~\cite{Gorghetto:2022sue}. If these structures rarely encounter the Solar System, then the effective value of the local DM density is substantially reduced. This highlights the importance of signals that instead rely on the DM density on Galactic scales, as investigated here. In future work, it would be interesting to more accurately estimate the sensitivity to such small dense DM substructures, as well as consider similar signals arising from new ultralight fields that do not make up the DM but are produced coherently during a supernova or constitute dark energy or dark radiation.

Compared to previously investigated signals of solar antineutrinos, the DM signal discussed here possesses unique  signatures. For instance, for \mbox{$\mAp \lesssim (v_\DM \, \AU)^{-1} \sim 10^{-15} \ \eV$}, each neutrino traverses the Sun-Earth distance in a background of DM with fixed phase. This phase randomly varies every time the Solar System encounters a new coherent region, corresponding to $t \sim \lambda_\DM / v_\DM \sim \text{week} \times (10^{-15} \ \eV / \mAp)$. On such timescales, the antineutrino flux should vary by $\order{1}$ factors.  Furthermore, for $(\g \, \text{AU})^{-1} \lesssim \mAp \lesssim \text{AU}^{-1}$, the elliptical polarization of the $\Ap$ field implies that a Solar antineutrino signal should have a burst-like structure, peaking every DM oscillation $\mAp^{-1} \gtrsim 10 \ \text{min}$ with a time-duration of $\text{AU} \sim 10 \ \text{min}$.  Although beyond the scope of this work, a dedicated analysis incorporating these effects may help reduce backgrounds and enhance sensitivity at significantly lower energy thresholds. 

Variations of the signals discussed here would also be worth exploring. For instance, DM-induced spin-precession of Dirac neutrinos flips chirality in addition to helicity, leading to neutrino disappearance instead of antineutrino appearance, analogous to signals of magnetic moments~\cite{Fujikawa:1980yx,Schechter:1981hw,Lim:1987tk, Akhmedov:1988uk} (see also Refs.~\cite{Kopp:2022cug,Alok:2022pdn,Jana:2022tsa,Brdar:2023cub,Sasaki:2023sza} for recent studies). CP-violation in the neutrino sector could manifest itself in effects tied to off-diagonal interactions of the $\Ap$ with neutrino mass-eigenstates. Finally, in dense astrophysical environments, neutrino helicity-flips could be resonantly amplified when the DM mass coincides with the in-medium energy-splitting.


\vspace{0.5cm}
\section*{Acknowledgments}

We thank Hannah Banks, John Beacom, Paddy Fox, Anson Hook, Ryan Plestid, and Cristina Volpe for valuable discussions. This manuscript has been authored by Fermi Forward Discovery Group, LLC under Contract No. 89243024CSC000002 with the U.S. Department of Energy, Office of Science, Office of High Energy Physics. This material is based upon work supported by the U.S. Department of Energy, Office of Science, National Quantum Information Science Research Centers, Superconducting Quantum Materials and Systems Center (SQMS) under contract number DE-AC02-07CH11359.
This work was partially supported by the University of Iowa's Year 2 P3 Strategic Initiatives Program through funding received for the project entitled ``High Impact Hiring Initiative (HIHI): A Program to Strategically Recruit and Retain Talented Faculty.''


\bibliography{ref}



\clearpage
\newpage
\maketitle
\onecolumngrid
\appendix
\begin{center}
\textbf{\large Neutrino-Antineutrino Conversion from Ultralight Vector Dark Matter} \\ 
\vspace{0.05in}
{ \it \large Supplemental Material}\\ 
\vspace{0.05in}
{}
{Asher Berlin, Rodolfo Capdevilla, Ting Cheng, Matheus Hostert, Pedro A. N. Machado}

\end{center}
\setcounter{equation}{0}
\setcounter{figure}{0}
\setcounter{table}{0}
\setcounter{section}{1}
\renewcommand{\theequation}{S\arabic{equation}}
\renewcommand{\thefigure}{S\arabic{figure}}
\renewcommand{\thetable}{S\arabic{table}}
\interfootnotelinepenalty=10000 

\section*{Neutrino Rest Frame Hamiltonian}
\label{app:QM}

In this section, we derive the low-energy Hamiltonian. Let us consider a single generation of neutrinos coupled to a vector field $\Apm$ as in \Eq{Lag1},  
\be
\label{eq:LagPsi1}
\Lag = \n^\dagger i \bar{\sigma}^\mu \partial_\mu \n - \Big( \frac{m_\n}{2} \, \n \n + \text{h.c.} \Big) + e^\p \Ap_\mu \, \n^\dagger \bar{\sigma}^\mu \n
~.
\ee
The classical equations of motion for $\n$ and $\n^\dagger$ are then
\be
\label{eq:PsiEOM1}
i \bar{\sigma}^\mu \partial_\mu \n - m_\n \, \n^\dagger + e^\p \Ap_\mu \, \bar{\sigma}^\mu \n = 0
~~,~~
i \sigma^\mu \partial_\mu \n^\dagger - m_\n \, \n - e^\p \Ap_\mu \, \sigma^\mu \n^\dagger = 0
~.
\ee
The first equation above determines $\n^\dagger$ in terms of $\n$,
\be
\n^\dagger = \frac{1}{m_\n} \, \big( i \bar{\sigma}^\mu \partial_\mu \n + e^\p \Ap_\mu \, \bar{\sigma}^\mu \n \big)
~.
\ee
This can then be used in the second equation of \Eq{PsiEOM1} to obtain an equation of motion solely in terms of $\n$,
\be
\label{eq:KG1}
( \partial^2 + m_\n^2 ) \, \n = i e^\p \, \sigma^\mu \bar{\sigma}^\n \, (\partial_\mu \Ap_\n) \, \n + i e^\p \, \Ap_\n \, \sigma^{[ \mu} \bar{\sigma}^{\n ]} \, \partial_\mu \n - e^{\p \, 2} A^{\p \, 2} \, \n
~.
\ee
Using the identities, 
\begin{align}
\sigma^\mu \bar{\sigma}^\n \, (\partial_\mu \Ap_\n) 
&= \partial_\mu \Apm + \sigmav \cdot (\grad A^{\p \, 0} + \dt \Av^\p + i \grad \times \Av^\p )
\nl
\Ap_\n \, \sigma^{[ \mu} \bar{\sigma}^{\n ]} \, \partial_\mu \n
&= 2 A^{\p \, 0} \, \sigmav \cdot \grad \n + 2 \Av^\p \hspace{-0.05cm} \cdot \sigmav \, \dt \n + 2 i \Av^\p \hspace{-0.05cm} \cdot (\sigmav \times \grad \n)
~,
\end{align}
\Eq{KG1} becomes
\be
(\partial^2 + m_\n^2) \, \n = i e^\p \, ( \grad A^{\p \, 0} + \dt \Av^\p + i \grad \times \Av^\p) \cdot \sigmav \, \n  + 2 i e^\p A^{\p \, 0} \, \sigmav \cdot \grad \n + 2 i e^\p \Av^\p \hspace{-0.05cm} \cdot \sigmav \, \dt \n - 2 e^\p \Av^\p \hspace{-0.05cm} \cdot (\sigmav \times \grad \n) - e^{\p \, 2} A^{\p \, 2} \, \n
~,
\ee
where we used that $\partial_\mu \Apm = 0$.

We now switch to a frame in which $\n$ is approximately at rest, by first factoring out the dominant part of $\n$'s time-evolution, $\n \equiv e^{-i m_\n t} \, \psi_\n$, such that $\psi_\n$ only includes the more slowly-oscillating behavior of the field,  $i \dt \psi_\n \ll m_\n \psi_\n$. Proceeding in this manner, we find
\begin{align}
\label{eq:Schrod1}
\Big( 1 + \frac{e^\p}{m_\n} \, \Av^\p \hspace{-0.05cm} \cdot \sigmav \Big) \, i \dt \psi_\n 
\simeq &- \frac{\grad^2 \psi_\n}{2 m_\n}
- \frac{i e^\p}{2 m_\n} \, ( \grad A^{\p \, 0} + \dt \Av^\p + i \grad \times \Av^\p) \cdot \sigmav \, \psi_\n  
\nl
&- \frac{i e^\p}{m_\n} \, A^{\p \, 0} \, \sigmav \cdot \grad \psi_\n 
- e^\p \Av^\p \hspace{-0.05cm} \cdot \sigmav \, \psi_\n 
+ \frac{e^\p}{m_\n} \Av^\p \hspace{-0.05cm} \cdot (\sigmav \times \grad \psi_\n) 
+ \frac{e^{\p \, 2}}{2 m_\n} A^{\p \, 2} \, \psi_\n
~.
\end{align}
The equation above is of the form $M \, i \dt \psi_\n = \tilde{H} \, \psi_\n$ with $M \equiv (1 + e^\p \Av^\p \hspace{-0.05cm} \cdot \sigmav / m_\n)$. This can be thought of as following from the effective Lagrangian \mbox{$\Lag = \psi_\n^\dag \, M i \dt \, \psi_\n - \psi_\n^\dagger \, \tilde{H} \, \psi_\n$}, which has a corresponding charge density $\psi_\n^\dag \, M \, \psi_\n$. Thus, to canonically normalize the field, we redefine $\psi_\n = M^{-1/2} \, \psi \simeq (1 - e^\p \Av^\p \hspace{-0.05cm} \cdot \sigmav / 2 m_\n) \, \psi$. In terms of this properly normalized $\psi$ field, \Eq{Schrod1} then becomes $i \dt \psi = H \psi$, where 
\be
\label{eq:Schrod2}
\eqbox{
H \simeq -\frac{\grad^2}{2 m_\n} - \frac{e^\p}{2 m_\n} \, \sigmav \cdot \big\{ i \grad \, , \, A^{\p \, 0} \big\} - e^\p \, \Av^\p \hspace{-0.05cm} \cdot \sigmav  + \frac{e^\p}{2 m_\n} \, \big\{ (\sigmav \times \grad)^i \, , \, A^{\p \, i} \big\}
}
\ee
is the low-energy Hamiltonian, we dropped terms of $\order{e^{\p \, 2}}$ and $\order{e^\p v^2}$ (with $v$ the neutrino velocity), and we used the operator identities
\begin{align}
\{ \grad , f(x) \} &= \grad f(x) + 2 f(x) \, \grad
\nl
(\sigmav \times \grad) \cdot \fv(x) &= \sigmav \cdot (\grad \times \fv(x)) + \fv(x) \cdot (\sigmav \times \grad)
~.
\end{align}
The second and third terms of \Eq{Schrod2} are the quantum mechanical analogues of the effective vector potential and effective magnetic field interactions that were previously identified in the classical result of \Eq{ClassicalLag1}, whereas the fourth term of \Eq{Schrod2}  is a relativistic correction to the third term.

\section*{Conversion Probability}
\label{app:elliptical}

Here, we derive an analytic approximation for the $\nnbar$ conversion probability. We start by considering the general interaction Hamiltonian for the neutrino spin in the $\nrf$,
\be
\label{eq:circH1}
H = V_x (t) \, \sigma_x +V_y (t) \, \sigma_y  + \big( V_z (t) + \Delta V_z \big) \, \sigma_z
~,
\ee
where $V_{x,y,z} (t)$ are dimensionful functions that parametrize the time-dependent interaction with the DM field and $\Delta V_z$ is a small constant offset that can arise from in-medium matter effects. In terms of $\sigma_\pm \equiv (\sigma_x \pm i \sigma_y) / 2$, \Eq{circH1} can be rewritten as
\be
\label{eq:circH2}
H =V_+ (t) \, \sigma_+ + V_- (t) \, \sigma_- + \big( V_z (t) + \Delta V_z \big) \, \sigma_z
~,
\ee
where we defined $V_\pm \equiv  V_x \mp i V_y$. We proceed by applying a unitary transformation $U = e^{- i \alpha_z \sigma_z}$, where $\alpha_z$ is some general function of time. Applying this transformation to \Eq{circH2}, the rotated Hamiltonian $H^\p = U^\dagger H \, U - i \, U^\dagger \dt U$ is
\be
\label{eq:circH3}
H^\p = V_+  \, e^{2 i \alpha_z} \, \sigma_+ + V_-  \, e^{-2 i \alpha_z} \, \sigma_- + \big( V_z + \Delta V_z - \dot{\alpha}_z \big) \, \sigma_z
~.
\ee
If we choose $\dot{\alpha}_z = V_z (t)$, then \Eq{circH3} becomes 
\be
\label{eq:circH4}
H^\p = V_+  \, e^{2 i \alpha_z} \, \sigma_+ + V_-  \, e^{-2 i \alpha_z} \, \sigma_- + \Delta V_z \, \sigma_z
~.
\ee
%

\subsection{Monochromatic Potential}

As a first toy example, let us take $V_{x,y,z}$ to be monochromatic, oscillating at one particular frequency $\w_0$, 
\be
\label{eq:Hcoherent1}
V_x = V_\perp \, \cos{(\w_0 t + \phi_x)}
~~,~~
V_y = V_\perp \, \cos{(\w_0 t + \phi_y)}
~~,~~
V_z = V_\parallel \, \cos{(\w_0 t)}
~,
\ee
where $V_\perp$ and $V_\parallel$ parametrize the strength of the DM interaction transverse or parallel to the neutrino's velocity, respectively. In the special case that $\phi_x = \phi_y = 0$, then the vector $\Vv \equiv (V_x , V_y , V_z)$ is linearly polarized. Instead, if, e.g., $\phi_x = 0$ and $\phi_y = - \pi / 2$, then $\Vv$ has a circular polarization component in the transverse plane. More generally, $\Vv$ traces out an ellipse in some plane. Note that \Eq{Hcoherent1} can be mapped onto the case of vector DM in the $\nrf$ by taking  
\be
\w_0 = \g \, \mAp
~~,~~
V_\perp = e^\p \, \frac{\sqrt{2 \rho_\DM / 3}}{\mAp}
~~,~~
V_\parallel = \g \, V_\perp
~~,~~
t = L / \g
~.
\ee
Using the form of the potential in \Eq{Hcoherent1}, we have that
\be
V_+ = \frac{V_\perp}{2} \, \big( c_+ \, e^{i \w_0 t} + c_- \, e^{-i \w_0 t} \big)
~~,~~
V_- = V_+^* 
~~,~~
\alpha_z = \frac{V_\parallel}{\w_0} \, \sin{(\w_0 t)}
~,
\ee
where we defined
\be
c_\pm \equiv e^{\pm i \phi_x} - i \, e^{\pm i \phi_y}
~.
\ee
With this form of $\alpha_z$, we can use the Jacobi-Anger expansion to rewrite the exponential factor in \Eq{circH4} as
\be
e^{2 i \alpha_z} 
= e^{i \, (2 V_\parallel / \w_0) \, \sin{(\w_0 t)}} 
=  \sum_{n = -\infty}^\infty  J_{n} \bigg( \frac{2 V_\parallel}{\w_0} \bigg) ~ e^{i n \w_0 t}
~,
\ee
which then gives 
\begin{align}
\label{eq:Hcoherent2}
H^\p &= \, \Delta V_z \, \sigma_z  + V_\perp \, J_1 (2 V_\parallel / \w_0) \, \big( \sin{\phi_x} \, \sigma_y - \sin{\phi_y} \, \sigma_x \big) 
\nl
&~~ + \Bigg( \frac{1}{2} \, V_\perp \sum_{n \neq \pm 1} J_n (2 V_\parallel / \w_0) \, \big(  c_+ \, e^{i (n+1) \w_0 t} + c_- \, e^{i (n-1) \w_0 t} \big) \, \sigma_+ + \text{h.c.} \Bigg)
~,
\end{align}
where $J_n$ is a Bessel function of the first kind, and we separated the static terms in the first line from the time-oscillating terms in the second line.

We expect the static terms in the first line of \Eq{Hcoherent2} to dominate the time-evolution, provided that the oscillation is sufficiently rapid, i.e., $\w_0 t \gg 1$, and that the  higher-order oscillating terms are smaller than the leading-order oscillating terms, which requires $\w_0 \gg V_\perp$. In this limit, we can thus approximate the time-evolution just with the first line of \Eq{Hcoherent2}, which we rewrite as $H^\p \simeq \Omegav \cdot \sigmav$, where 
\be
\Omegav \equiv \big( -V_\perp \, \sin{\phi_y} \, J_1 (2 V_\parallel / \w_0) ~,~ V_\perp \, \sin{\phi_x} \, J_1 (2 V_\parallel / \w_0) ~,~ \Delta V_z \big)
~.
\ee
Writing $\Omegav = \Omega \, \hat{\Omegav}$ with $\hat{\Omegav}$ a unit-vector, the time-evolution operator is approximately $U(t) \simeq \cos{(\Omega \, t)} - i \, \sin{(\Omega \, t)} ~ \hat{\Omegav} \cdot \sigmav$. 
This therefore gives a spin-flip probability of
\be
\label{eq:Prcoherent0}
\Prn \simeq | \upbra U(t)\downket |^2 \simeq  \hat{\Omega}_\perp^{\, 2} \, \sin^2{(\Omega \, t)}
~,
\ee
where the up and down arrows refer to spin along the $\pm z$ directions, respectively, and $\hat{\Omega}_\perp^{\, 2} = \hat{\Omega}_x^{\, 2}+ \hat{\Omega}_y^{\, 2}$. Note that this vanishes in the $\phi_x , \phi_y \to 0$ (linearly-polarized) limit. In the limit that $\Delta V_z \ll V_\perp \, J_1(2 V_\parallel / \w_0)$, \Eq{Prcoherent0} becomes
\be
\label{eq:Prcoherent1}
\Prn \simeq  \sin^2{\Big( \sqrt{\sin^2{\phi_x} + \sin^2{\phi_y}} ~ V_\perp \, J_1(2 V_\parallel / \w_0) \, t \Big)}
~,
\ee
whereas for  $\Delta V_z \gg V_\perp \, J_1(2 V_\parallel / \w_0)$ we have
\be
\label{eq:Prcoherent1b}
\Prn \simeq \frac{ \big( \sin^2{\phi_x} + \sin^2{\phi_y} \big) \, V_\perp^2 \, J_1^2(2 V_\parallel / \w_0)}{\Delta V_z^2} \, \sin^2{(\Delta V_z \, t)}
~.
\ee
This latter case shows that when $\Delta V_z$ is large, the maximum value of $\Prn$ is much less than unity. This is to be expected, since at the level of standard Rabi oscillations a static potential is important when $\Delta V_z \gg V_\perp$~\cite{Sakurai:2011zz}. However, note that this is not evident to leading order in perturbation theory, since \Eqs{Prcoherent1}{Prcoherent1b} take the same form for $V_\perp \, ,\,  \Delta V_z \to 0$.

In the weak-coupling limit, \Eqs{Prcoherent1}{Prcoherent1b} scale quartically in the potential, $\Prn \propto V_\perp^2 \, V_\parallel^2$. This is due to the fact that the quadratic contribution does not lead to a conversion probability that continues to grow with time $t \gg \w_0^{-1}$. In the derivation above, this can be seen from the fact that both $V_\perp$ and $V_\parallel$ need to be non-zero to yield a static term in the rotated Hamiltonian of \Eq{Hcoherent2}. Although not shown here, this can alternatively be made more explicit by working within the original basis of \Eq{circH1} and demonstrating that the quadratic terms yield a conversion probability that is bounded as $\Prn \lesssim V_\perp^2 / \w_0^2$ at late times.

\subsection{Non-Monochromatic Potential}

The above example took the potential to oscillate coherently for arbitrarily long times. As a result, the probability in \Eq{Prcoherent1} scales as $\Prn \propto t^2$. However, for times much longer than some coherence timescale $\Delta t_\text{coh}$, we expect the scaling to soften to $\Prn \propto t$. We now determine the specific form of a finite coherence time coming from the non-zero DM dispersion velocity $v_\DM$. In particular, we consider an example similar to \Eq{Hcoherent1}, but where $V_z$ has a new term that oscillates at a slightly different frequency than the others, 
\be
\label{eq:Hdrift1}
V_x = V_\perp \, \cos{(\w_0 t + \phi_x)}
~~,~~
V_y = V_\perp \, \cos{(\w_0 t + \phi_y)}
~~,~~
V_z = \frac{V_\parallel}{2} \, \Big( \cos{(\w_0 t)} +  \cos{\big( (\w_0 + \Delta \w) t \big)} \Big)
~.
\ee
In the expression for $V_z$ above, $\Delta \w \ll \w_0$ is a small frequency shift. In principle, there are analogous terms for the transverse components $V_{x,y}$, but these have a subdominant effect. To map this onto the problem of a relativistic neutrino propagating through a DM background, $2 \pi / \Delta \w$ is related to the time it takes to traverse a DM coherence length after boosting into the $\nrf$, such that
\be
\label{eq:freqdrift1}
\Delta \w \simeq \w_0 \, v_\DM \simeq \g \, \mAp \, v_\DM
~.
\ee

The resulting form for $V_\pm$ is identical to the previous example. However, $\Delta \w \neq 0$ modifies $\alpha_z$. In the limit that $\w_0^{-1} \ll t \ll \Delta \w^{-1}$, this is given by
\be
\label{eq:alphadrift}
\alpha_z \simeq V_\parallel \, \bigg( \frac{1}{\w_0} \, \sin{(\w_0 t)} + \frac{\Delta \w}{2 \w_0} \, t \, \cos{(\w_0 t)} \bigg)
~.
\ee
Once again, we can use the Jacobi-Anger expansion to rewrite the exponential factor in \Eq{circH4} as
\begin{align}
e^{2 i \alpha_z} &\simeq e^{i \, (2 V_\parallel / \w_0) \, \sin{(\w_0 t)}} \, e^{i \, (\Delta \w / \w_0) \, V_\parallel t \, \cos{(\w_0 t)}} 
=  \sum_{n_1, n_2 = -\infty}^\infty  \hspace{-0.25cm} J_{n_1} \bigg( \frac{2 V_\parallel}{\w_0}  \bigg) ~  J_{n_2} \bigg( \frac{\Delta \w}{\w_0} \, V_\parallel \, t \bigg) ~ e^{i (n_1 + n_2) \w_0 t} ~ e^{i \pi n_2   /2}
~,
\end{align}
such that the rotated Hamiltonian is
\be
\label{eq:Hdrift2}
H^\p = \Delta V_z \, \sigma_z + \Bigg( \sigma_+ \, \frac{V_\perp}{2} \, \big( c_+ \, e^{i \w_0 t} + c_- \, e^{-i \w_0 t} \big)  \hspace{-0.25cm} \sum_{n_1, n_2 = -\infty}^\infty  \hspace{-0.25cm} J_{n_1} \bigg( \frac{2 V_\parallel}{\w_0}  \bigg) ~  J_{n_2} \bigg( \frac{\Delta \w}{\w_0} \, V_\parallel \, t \bigg) ~ e^{i (n_1 + n_2) \w_0 t} ~ e^{i \pi n_2   /2} ~  + \text{h.c.} \Bigg)
~.
\ee
This shows that when $t \ll \min{(1 \, , \, \, \w_0 / V_\parallel}) \, \Delta \w^{-1}$, the frequency drift is negligible and we recover the same result as in the case of a monochromatic potential. On the other hand, the frequency drift is important when $t \gg \min{(1 \, , \, \, \w_0 / V_\parallel}) \, \Delta \w^{-1}$, which occurs well before $t \sim \Delta \w^{-1}$ if $\w_0 \ll V_\parallel$. In this limit, the non-oscillating terms involving $\sigma_\pm$ in \Eq{Hdrift2} are power-law suppressed for large $t$, due to $J_n (x) \propto 1/\sqrt{x}$ for $x \gg 1$. Since there are no unsuppressed non-oscillating $\sigma_\pm$ terms, the Jacobi-Anger expansion is not particularly useful. 

To tackle the regime  $t \gg \min{(1 \, , \, \, \w_0 / V_\parallel}) \, \Delta \w^{-1}$, we instead revisit the general form of the potential in \Eq{circH4}. To allow for general time-dependence of the Hamiltonian, let us absorb $\Delta V_z$ into the definition of $V_z (t)$, such that \Eq{circH1} is instead $H = V_x (t) \, \sigma_x +V_y (t) \, \sigma_y  + V_z (t) \, \sigma_z$ and the analogous form of
\Eq{circH4} is
\be
\label{eq:circH4alt}
H^\p = V_+  \, e^{2 i \alpha_z} \, \sigma_+ + V_-  \, e^{-2 i \alpha_z} \, \sigma_- 
~.
\ee
In general, the time-evolution operator can be approximated to leading order as $U(t) \simeq 1 -  i \int_0^t dt^\p ~ H^\p (t^\p)$, which gives for the transition amplitude
\be
\upbra U(t)\downket \simeq -i \,  \int_0^t dt^\p ~ \upbra \,H^\p (t^\p) \,\downket
= - i \,  \int_0^t dt^\p ~  V_- (t^\p)  \, e^{-2 i \alpha_z(t^\p)} 
~.
\ee
To evaluate the time-integral, we utilize a saddle-point approximation. In particular, if $e^{-2 i \alpha_z (t)}$ is a rapidly oscillating function of time, corresponding to $V_\parallel  \, \min{(\w_0^{-1} , t)} \gg 1$, we can expand around the saddle-point(s) $t_0$, defined via $\dot{\alpha}_z(t_0) = V_z (t_0) = 0$, such that $\alpha_z (t) \simeq  \alpha_z (t_0) +  \frac{1}{2} \, (t - t_0)^2 \, \dot{V}_z (t_0)$. Therefore, summing over multiple possible saddle-points $t_0$ and approximating the time-integral as spanning the entire domain, we have the approximate form
\be
\upbra U(t)\downket \simeq  i \, \sum_{t_0} e^{-2 i  \alpha_z (t_0)}  \, V_- (t_0)  \, \int_{-\infty}^\infty d\tau ~ e^{- i \tau^2 \, \dot{V}_z (t_0)}  
~.
\ee
Evaluating the integral gives
\be
\label{eq:ampgen1}
\upbra U(t)\downket \simeq  - i \, \sum_{t_0} e^{-2 i  \alpha_z (t_0)}  \, e^{- i \,  (\pi/4) \, \text{sign}[\dot{V}_z (t_0)]} ~ V_- (t_0)  \, \sqrt{\frac{\pi}{|\dot{V}_z (t_0)|}}
~.
\ee
Taking $V_- \sim V_\perp$, $V_z \sim V_\parallel$, $\dot{V}_z \sim \w_0 \, V_\parallel$, the various levels of approximation can therefore be summarized as: ``large phase" $\alpha_z \sim \min{(\w_0^{-1} , t)} \, V_\parallel \gg 1$, ``perturbativity" \mbox{$V_\perp \ll \sqrt{\w_0 \, V_\parallel}$}, and ``long-time" $t \gg 1 / \sqrt{\w_0 \, V_\parallel}$, corresponding to
\be
V_\parallel \gg \max{\bigg( \frac{V_\perp^2}{\w_0} \, , \, \w_0 \, , \, \frac{1}{t} \, , \, \frac{1}{\w_0 \, t^2} \bigg)}
~.
\ee

As a cross-check, we apply \Eq{ampgen1} to the previously investigated case of a monochromatic potential of \Eq{Hcoherent1} (but with our shifted definition of $V_z$),
\be
V_x = V_\perp \, \cos{(\w_0 t + \phi_x)}
~~,~~
V_y = V_\perp \, \cos{(\w_0 t + \phi_y)}
~~,~~
V_z = V_\parallel \, \cos{(\w_0 t)} + \Delta V_z
~.
\ee
The saddle-points are defined by $V_z(t_0) = 0$. In the limit that $\Delta V_z \ll V_\parallel$, these are given by
\be
\label{eq:t0saddle}
t_0 \simeq \frac{1}{\w_0} \, \bigg[ 2 \pi n \pm \bigg( \frac{\pi}{2} + \frac{\Delta V_z}{V_\parallel} \bigg) \bigg]
~.
\ee
Note that $\Delta V_z$ enters the phase $\alpha_z$ via $\alpha_z = (V_\parallel / \w_0) \, \sin{(\w_0 t)} + \Delta V_z \, t$. Therefore, \Eq{ampgen1} implies that saddle-points that are separated in time by $\gtrsim 1/\Delta V_z$ do not add coherently. For this example, the sum in \Eq{ampgen1} can be evaluated analytically from $n = 0$ to $n=\w_0 t / 2\pi$, including both of the $\pm$ branches of \Eq{t0saddle}. In particular, we find that in the limit $\Delta V_z \ll V_\parallel \, , \, \w_0$, \Eq{ampgen1} gives
\be
\label{eq:ampcoh1}
\upbra U(t)\downket \simeq  - e^{-i \Delta V_z t} \, \big( \sin{\phi_x} + i \sin{\phi_y} \big) \, V_\perp \, \sqrt{\frac{\w_0}{ \pi \, V_\parallel}} \, \sin{(\pi / 4 - 2 V_\parallel / \w_0)} \,  \frac{\sin{(\Delta V_z (t + 2 \pi / \w_0))}}{ \Delta V_z} 
~.
\ee
After squaring, this is consistent with our previous derivation of \Eqs{Prcoherent1}{Prcoherent1b}, after taking  $V_\parallel \gg \w_0$.

In the absence of matter effects ($\Delta V_z \to 0$), the amplitude in \Eq{ampcoh1} scales linearly with time because the potential was taken to be monochromatic, and hence coherent across the time interval. However, this is only true for groups of saddle-points $t_0$ for which $\alpha_z (t_0)$ does not evolve by more than $\order{1}$. From the discussion near \Eq{Hdrift2}, this occurs on timescales 
\be
\label{eq:DeltatCoh1}
\Delta t_\text{coh} \simeq 2\pi \, \Delta \w^{-1} \, \min(1 \, , \, \w_0 / V_\parallel)
~,
\ee
where $\w_0$ is the characteristic frequency of the potential, and $\Delta \w \ll \w_0$ is the frequency spread of the potential defined in \Eq{freqdrift1}. Thus, the parametrics of \Eq{ampcoh1} can be understood by noting that for $\w_0^{-1} \ll t \ll \Delta t_\text{coh}$, each saddle-point in \Eq{ampgen1} adds coherently  (which occurs every oscillation time $2 \pi / \w_0$), such that 
\be
\Prn \simeq  \bigg[ \bigg( \frac{\w_0 \, t}{2 \pi} \bigg) \, \bigg( V_\perp \, \sqrt{\frac{\pi}{\w_0 \, V_\parallel}} \bigg) \bigg]^2 
= 
\bigg( \frac{V_\perp^2}{V_\parallel} \, \w_0 \, t \bigg) 
\, \frac{t}{4\pi}
~,
\ee
which reproduces the square of \Eq{ampcoh1} up to an $\order{1}$ factor. 

Next, we consider the case that $\w_0^{-1} \ll \Delta t_\text{coh} \ll t$, such that within each interval of duration $\Delta t_\text{coh}$, there are $\sim \w_0 \, \Delta t_\text{coh} / 2 \pi$ saddle-points that add coherently to the sum of \Eq{ampgen1}, and each of the $t / \Delta t_\text{coh}$ groupings of such saddle-points add incoherently,
\be
\Prn \simeq  \bigg[ \sqrt{\frac{t}{\Delta t_\text{coh}}} \, \bigg( \frac{\w_0 \, \Delta t_\text{coh}}{2 \pi} \bigg) \, \bigg( V_\perp \, \sqrt{\frac{\pi}{\w_0 \, V_\parallel}} \bigg) \bigg]^2 
= 
\bigg(\frac{V_\perp^2}{V_\parallel} \, \w_0 \, t \bigg) \, \frac{\Delta t_\text{coh}}{4 \pi}
~.
\ee
Instead, if $\Delta t_\text{coh} \ll \w_0^{-1} \ll t$, each saddle-point adds incoherently,
\be
\Prn \simeq  \bigg[ \sqrt{\frac{\w_0 \, t}{2 \pi}} \, \bigg( V_\perp \, \sqrt{\frac{\pi}{\w_0 \, V_\parallel}} \bigg) \bigg]^2 
= 
\bigg( \frac{V_\perp^2}{V_\parallel} \, \w_0 \, t \bigg) \, \frac{1}{2 \w_0}
~.
\ee
Finally, for $t \ll \w_0^{-1}$ there is only one saddle-point. However, in this case we need to include an additional factor corresponding to the probability $\sim \w_0 t / 2 \pi \ll 1$ that we pass through this saddle-point during the short time-interval,
\be
\Prn \simeq  \bigg( V_\perp \, \sqrt{\frac{\pi}{\w_0 \, V_\parallel}} \bigg)^2 \, \frac{\w_0 \, t}{2 \pi}
= 
\bigg( \frac{V_\perp^2}{V_\parallel} \, \w_0 \, t \bigg) \, \frac{1}{2 \w_0}
~,
\ee
which yields the same result as the previous example. All of these scenarios can be summarized simply as
\be
\label{eq:Princoherent1}
\Prn \simeq \w_0 \, t \, \frac{V_\perp^2}{2 V_\parallel}   ~ \text{max}\bigg[\frac{1}{\w_0} \, ,\, \min{\bigg(\frac{t}{2 \pi} \, , \,  \frac{\w_0}{V_\parallel \, \Delta \w} \bigg)}\bigg]
~,
\ee
where we took $V_\parallel \gg \w_0$. 

Finally, we note that from the discussion above, matter effects in the form of $\Delta V_z \neq 0$ lead to a dephasing in $\alpha_z$ in a way that is similar to the effect of the frequency spread of \Eq{DeltatCoh1}. Thus, we can ignore the role of $\Delta V_z$ if either $\Delta V_z \ll V_\perp$ or $\Delta V_z \ll \Delta \w \, \max(1 \, , \, V_\parallel / \w_0)$. This is satisfied throughout the parameter space of interest.

\subsection{Summary}

Let us briefly recap how we implement the above results. Throughout, we take
\be
\w_0 = \g \, \mAp
~~,~~
\Delta \w \simeq \w_0 \, v_\DM 
~~,~~
V_\perp = e^\p \, \frac{\sqrt{2 \rho_\DM / 3}}{\mAp}
~~,~~
V_\parallel = \g \, V_\perp
~~,~~
t = L / \g
~~,~~
\Delta t_\text{coh} \simeq \frac{2\pi}{\Delta \w} \, \min \bigg(1 \, , \, \frac{\w_0}{V_\parallel} \bigg)
~.
\ee
In the limit that $\w_0 \gg V_\parallel$, the coherence time is dictated solely by the DM coherence, i.e., $\Delta t_\text{coh} \simeq 2 \pi \, \Delta \w^{-1}$. We then use \Eq{Prcoherent1}, scaled linearly in time for $t \gg \Delta t_\text{coh}$. Instead, if $\w_0 \ll V_\parallel$, $\Delta t_\text{coh} \simeq 2 \pi \, \Delta \w^{-1} \, (\w_0 / V_\parallel)$, and we use \Eq{Princoherent1}.
This is summarized as
\begin{empheq}[box=\fcolorbox{light-gray}{light-gray}]{align}
\label{eq:P_summary}
\lim_{\w_0 \gg V_\parallel} \Prn &\simeq \max\bigg( 1 \, , \, \frac{t \, \Delta \w}{2 \pi} \bigg) \, \sin^2{\Bigg( V_\perp \, J_1 \bigg(\frac{2 V_\parallel}{\w_0} \bigg) \, \min{(t \, , \, 2 \pi \, \Delta \w^{-1})}  \Bigg)}
\nl
\lim_{\w_0 \ll V_\parallel} \Prn &\simeq
\w_0 \, t \, \frac{V_\perp^2}{2 V_\parallel}   ~ \text{max}\bigg[\frac{1}{\w_0} \, ,\, \min{\bigg(\frac{t}{2 \pi} \, , \,  \frac{\w_0}{V_\parallel \, \Delta \w} \bigg)}\bigg]
~,
\end{empheq}
where in \Eq{Prcoherent1} we replaced $\sqrt{\sin^2{\phi_x} + \sin^2{\phi_y}} \to 1$ with its RMS value.


\subsection{Numerical}
\label{app:numerical}

To confirm our analytic results above, we must solve the time-evolution numerically. We take the Hamiltonian for a neutrino traveling along the $+z$ direction to be $H = \Vv (t) \cdot \sigmav$, where $\Vv$ is a time-dependent three-vector. To represent an interaction with ultralight DM, we numerically evaluate each component of $\Vv$ as~\cite{Foster:2017hbq}
\be
\label{eq:DMfield2}
V^i (t) \simeq c^i \, \frac{V_\DM}{\sqrt{\mathcal{N}}} ~ \sum_{j = 1}^{\mathcal{N}} \cos{\big((\w_0 + \Delta \w_j) \, t  + \phi_j\big)}  
~,
\ee
where $V_\DM$ sets the strength of the interaction, and $\phi_j \in [0 , 2 \pi]$ and $\Delta \w_j \in [0 , \Delta \w]$ are uniformly distributed random variables, independently drawn when evaluating each component $V^i$. This expression is motivated by the fact that the DM field is composed of many particles $\mathcal{N} \gg 1$ drawn from a narrow velocity distribution.   In \Eq{DMfield2}, the coefficient $c^i$ accounts for the Lorentz boost of the longitudinal component of the $\Av^\p$ field in the $\nrf$, such that $c^{x,y} = 1$ and $c^z = \g$. Since each phase $\phi_j$ is random, the terms in the sum add incoherently, such that the characteristic magnitude of \Eq{DMfield2} is $|V^i| \sim c^i \, V_\DM$. To make contact with our notation from the above derivations, we take $V_\parallel \equiv \g \, V_\DM$, $V_\perp \equiv V_\DM$, and $\Delta \w / \w_0 \sim v_\DM \sim 10^{-3}$.

\begin{figure}[t]
\centering
\includegraphics[width=0.8 \textwidth]{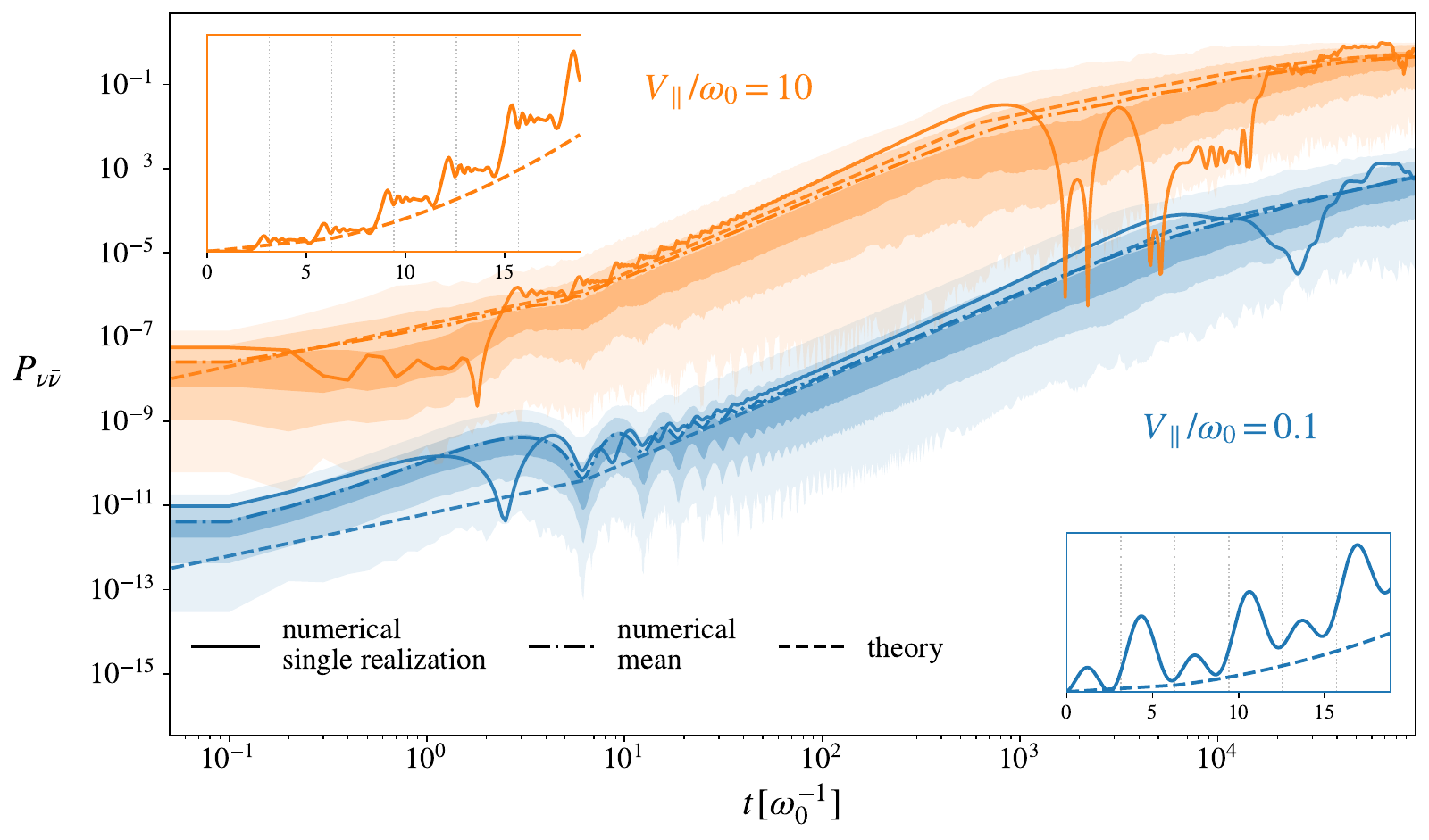}
\caption{A comparison of our analytic results to a numerical determination of the neutrino-antineutrino conversion probability $\Prn$, as a function of time $t$ in the neutrino rest frame (in units of $\w_0^{-1}$). We show the numerical output for either a single realization of the DM field as in \Eq{DMfield2} with $\mathcal{N} = 500$ modes (solid lines) or the mean of 500 such DM realizations (dot-dashed lines). For comparison, our analytic expressions from \Eq{P_summary} are also shown (dashed lines). 
The blue lines correspond to $V_\parallel/\w_0 = 0.1$ and the orange lines correspond to $V_\parallel/\w_0 = 10$. Throughout, we have fixed the neutrino boost to $V_\parallel/V_\perp =\gamma = 10^4$. The outer/middle/inner colored bands highlight the (1-99\%)/(10-90\%)/(32-68\%) quantile ranges of the data. 
The insets show a zoomed-in time-range with linear axes, with vertical grid lines spaced every $\pi / \w_0$ time-interval.
}
\label{fig:numerical}
\end{figure}

We next divide up the total time $t$ into small time-steps $t_n = n \, \Delta t$ with $\Delta t \ll \min{\big( 1 / V_\parallel \, , \, 1 / \w_0 \big)} \ll t$. In this case, the time-evolution operator can be numerically approximated as a product of constant time-evolutions,
\be
U(t) \simeq \prod_{n = 1}^{t / \Delta t} \bigg(  \cos{\big(|\Vv (t_n)| \, \Delta t\big)} - i \, \sin{\big(|\Vv (t_n)|\,  \Delta t\big)} \, \frac{\Vv (t_n) \cdot \sigmav}{|\Vv(t_n)|}  \bigg)
~.
\ee
In \Fig{numerical}, we use this to numerically compute the conversion probability $\Prn = |\bra{\,\uparrow}\, U(t)\, \ket{\downarrow\,}|^2$ for a neutrino to flip from spin along $-z$ to spin along $+z$, and compare this result to the previous analytical result of \Eq{P_summary}. This is shown for a large or small representative value of $V_\parallel / \w_0$. In each case, we perform $500$ random realizations of the DM field, with each realization built from $\mathcal{N} = 500$ modes drawn according to \Eq{DMfield2}. The numerical output of a single DM realization is shown in the solid lines. The ensemble-mean from many DM realizations is shown as the dot-dashed lines. Finally, the analytic expectation is shown as the dashed lines. This illustrates good agreement between our analytic expressions and numerical output.

The two insets of \Fig{numerical} show a zoomed-in linear time-axis spanning only a few DM oscillation periods, over which the DM field is approximately monochromatic. As discussed earlier near \Eq{ampgen1}, the dynamics are dominated by the saddle-points of the rotated Hamiltonian, which occur every DM oscillation time $\sim \w_0^{-1}$ when $\dot{\alpha}_z = V_z = 0$ and the two instantaneous energy levels become nearly-degenerate. In the non-adiabatic limit, $V_\perp^2 / (\w_0\, V_\parallel) \ll 1$, each such crossing transfers only a small amount of probability to the flipped state and can be treated perturbatively, as in \Eq{ampgen1}. 
Between successive crossings, the relative phase between the two energy levels evolves at a rate set by $V_\parallel$, so that contributions from different crossings interfere coherently. 
This is the familiar Landau--Zener--St\"uckelberg--Majorana interference of a periodically driven two-level system across avoided level crossings~\cite{Landau:1932,Zener:1932,Stuckelberg:1932,Majorana:1932}. 
For $V_\parallel/\w_0 \gg 1$ (orange), the crossings occur on timescales much longer than the intermediate phase evolution, such that the discrete jumps at each crossing dominate the conversion probability. As a result, for $t \ll \w_0^{-1}$, the transition probability is controlled by whether a single jump has yet occurred, giving $\Prn \propto t$, as displayed by the early-time behavior of the ensemble-mean (dot-dashed). For $V_\parallel/\w_0 \ll 1$ (blue line), these jumps occur on much shorter timescales than the oscillations, and hence are no longer clearly separated.

In either regime, the DM field drives a coherent build-up $\Prn \propto t^2$ for $\w_0^{-1} \ll t \ll \Delta t_\text{coh}$, due to the static term in the rotated Hamiltonian (\Eq{Prcoherent0}). This coherent build-up persists only as long as the main driving term in the Hamiltonian remains monochromatic, corresponding to $t \lesssim \Delta t_\text{coh}$.  This timescale varies between DM realizations, but its expectation value is captured by the ensemble-mean (dot-dashed lines) and computed analytically in \Eq{DeltatCoh1}. As expected from our analytics, the $\Delta t_\text{coh}$ differs in the two regimes $V_\parallel/\w_0 < 1$ and $V_\parallel/\w_0 > 1$. Once $t \gtrsim \Delta t_\text{coh}$, a single DM realization corresponds to stochastic evolution of $\Prn$ (solid), scaling on average as $\Prn \propto t$ (dashed). The colored bands exhibit non-Gaussian behavior, with the ensemble-mean offset from the median.

%

\section*{QFT Amplitude}
\label{app:QFT}

In the discussion above, we calculated the probability for a neutrino to flip its helicity. In the next two sections, we show explicitly at the level of quantum field theory that this also corresponds to $\nnbar$ conversion. We first consider a simpler problem than the one discussed above, and calculate a scattering amplitude. In particular, take the neutrino to couple to a static external potential with a Hamiltonian of the form $H = \Vv \cdot \sigmav$, where $\Vv = |\Vv| \, \hat{\Vv}$ is a constant vector. In this case, the time-evolution operator is trivial, such that the quantum mechanical probability to flip helicity is given by
\be
\label{eq:ProbQM}
\Prn \simeq | \upbra e^{-i \Vv \cdot \sigmav \, L / \g}\downket |^2 = \bigg( \frac{V_\perp}{|\Vv|} \bigg)^2 \, \sin^2{(|\Vv| L / \g)}
~,
\ee
where $V_\perp \equiv \sqrt{V_x^2 + V_y^2}$.

We can alternatively compute the Feynman diagram amplitude for $\n_{\downarrow} \to \n_{\uparrow}$, where the subscript denotes the direction of the spin and we take the neutrino to be traveling along the $+z$ direction. We first rewrite the Lagrangian corresponding to this interaction in four-component notation, $\Lag \supset -\frac{1}{2} \, V_\mu \, \bar{\Psi} \g^\mu \g^5 \Psi$, where $\Psi$ is the neutrino Majorana spinor. To write down the amplitude for $\n_{\downarrow} \to \n_{\uparrow}$, we follow the conventions of Ref.~\cite{Srednicki:2007qs}, which gives
\be
i \mathcal{M} 
\simeq - i \,  \bar{u}_f (p_\n) \, (V_x \, \g^1 + V_y \, \g^2) \, \g^5 u_i(p_\n)
~.
\ee
Here, $u_i (p_\n)$ is the four-spinor for the initial state, corresponding to a negative-helicity particle traveling in the $+z$ direction, whereas $u_f(p_\n)$ is the spinor for the final state, corresponding to a positive-helicity particle traveling in the $+z$ direction, i.e.,
\be
\label{eq:4spinors}
u_i(p_\n) = \mat{\sqrt{E_\n+p_\n} ~\downket \vspace{0.2cm} \\ \sqrt{E_\n-p_\n} ~\downket}
~~,~~
u_f(p_\n) = \mat{\sqrt{E_\n-p_\n} ~\upket \vspace{0.2cm} \\ \sqrt{E_\n+p_\n} ~\upket}
~,
\ee
which yields $i \mathcal{M} = 2 i \, m_\n \, (V_x - i V_y)$. The differential rate is then given by
\be
\frac{d \Prn}{dL} = \frac{1}{2 E_\n} \, |\mathcal{M}|^2 \, \text{dLIPS}_1 
~,
\ee
where the one-body Lorentz invariant phase space is $\text{dLIPS}_1 \simeq L / E_\n$. Thus, we have that the total rate is \mbox{$\Prn \simeq  \big( V_\perp \, L / \g \big)^2$}, which is in agreement with \Eq{ProbQM} to leading order in $|\Vv|$. Finally, note that the final-state positive-helicity Majorana neutrino will propagate with both positive- and negative-energy solutions, corresponding to the standard $u(p_\n)$ and $v(p_\n)$ spinors. The part of this final-state field that can interact with a detector (dominantly left-handed chirality) corresponds to the four-spinor component
\be
 v(p_\n) = \mat{\sqrt{E_\n+p_\n} ~\downket \vspace{0.2cm} \\ - \sqrt{E_\n-p_\n} ~\downket}
 ~,
\ee
which in the relativistic limit is equivalent to an antineutrino.

\section*{QFT Hamiltonian}
\label{app:Helicity}

Alternatively, we can show that the low-energy Hamiltonian used throughout this work can be derived explicitly from second-quantization of the neutrino field, where the Hilbert space explicitly corresponds to the $\n$--$\nbar$ system. To this end, let us reparameterize the interaction Lagrangian in four-component notation as $\Lag \supset -\frac{1}{2} \, V_\mu \, \bar{\Psi} \g^\mu \g^5 \Psi \equiv - V_\mu \, S^\mu$, where $\Psi$ is the Majorana neutrino field and $V^\mu \equiv e^\p \, A^{\p \, \mu}$ in the lab-frame. As discussed above, for a non-relativistic DM background, we can neglect $A^{\p \, 0}$ in the lab-frame, such that the lab-frame Hamiltonian is
\be 
\label{eq:SemiClassicalH_lab}
H_\text{lab} \simeq -e^\p \, \Av^\p \cdot \Sv
~.
\ee
Following the conventions of Ref.~\cite{Srednicki:2007qs}, $\Psi$ can be expanded in terms of creation and annihilation operators $a^{\dagger}_\pm$ and $a_\pm$ and positive- and negative-energy four-spinors in the helicity ($\pm$) basis,
\begin{align}
\label{eq:spinors}
&u_{+} =    
\mat{
\sqrt{E_\n-p_\n} ~ \xi_+ \\
\sqrt{E_\n+p_\n} ~ \xi_+
}
~~,~~
u_{-} =    
\mat{
\sqrt{E_\n+p_\n} ~ \xi_- \\
\sqrt{E_\n-p_\n} ~ \xi_-
} 
~~,~~
v_{+} =
\mat{
\sqrt{E_\n+p_\n}  ~ \xi_-\\
-\sqrt{E_\n-p_\n} ~ \xi_-
}
~~,~~
v_{-} =
\mat{
-\sqrt{E_\n-p_\n} ~ \xi_+\\
\sqrt{E_\n+p_\n}  ~ \xi_+
}
~,
\end{align} 
where $\left(\sigmav \cdot \hat{\vp}_\n \right)\, \xi_\pm=\pm \, \xi_\pm$. We then perform this expansion in $S^\mu$, and map the four-component spinors onto a two-level system via $|\n\rangle =  a^\dagger_- |0\rangle$ and $|\nbar\rangle  = a^\dagger_+ |0\rangle$, which makes contact with the observational definitions of neutrinos and antineutrinos. Since under charge-conjugation $\Psi = \Psi^C = \mathcal{C} \,  \bar\Psi^{T}$ for Majorana neutrinos, $S^i$ takes the form
\be
S^i\simeq 
\xi^\dagger_- \sigma^i \xi_- \, |\n \rangle\langle\n|
+ \frac{1}{\g} \, \xi^\dagger_- \sigma^i \xi_+\, |\n \rangle\langle\nbar|
+ \frac{1}{\g} \, \xi^\dagger_+\sigma^i\xi_-\, |\nbar\rangle\langle\n|
+\xi^\dagger_+\sigma^i \xi_+\, |\nbar \rangle\langle\nbar|
~.
\ee
Finally, using the identities $\big(\mathbf{A}^\p \cdot \sigmav \big)\, \xi_{\pm} = \pm A'_\parallel \, \xi_\pm+A'_\perp e^{\pm i\phi} \, \xi_\mp$ where $A'_1-iA'_2 \equiv A'_\perp \, e^{i\phi}$, the lab-frame Hamiltonian in the $\n$--$\nbar$ basis becomes
\be
H_\text{lab} \simeq e^\p \, A^\p_\parallel \, \sigma_z +  \frac{1}{\g}\, e^\p \, A'_\perp 
\left(e^{-i\phi}\sigma_+ +  e^{i\phi}\sigma_- \right)
~,
\ee
which is equivalent to the Hamiltonian used throughout this work after boosting to the lab-frame. 

\end{document}